\documentstyle[preprint]{jpsj}

\def\runtitle{
Conductance in 
Quantum Wires
}
\def\runauthor{S. Fujimoto and N. Kawakami}

\title{Effect of Umklapp Scattering on the Conductance \\
in Quantum Wires}

\author{Satoshi Fujimoto$^1$ and Norio Kawakami$^2$}
\inst{$^1$Department of Physics, Kyoto University, Kyoto 606, Japan\\
$^2$Department of Applied Physics,
and Department of Material and Life Science, \\
Osaka University, Suita, Osaka 565, Japan}

\recdate{}
  
\abst{
We investigate the system-size dependence
as well as the temperature dependence of the
conductance in 1D electron systems, paying particular
attention to the effect of Umklapp scattering.
By taking into account the renormalization of
the external potential due to electron-electron interaction,
the correction to the conductance, $2e^2/h$, 
due to Umklapp scattering is estimated perturbatively.
The Tomonaga-Luttinger liquid parameter appears
in the conductance via the system-size and temperature 
dependence.
}
\kword
{conductance, Tomonaga-Luttinger liquid, quantum wire}
\begin{document}
\sloppy
\maketitle

The effect of electron-electron interaction on 
the conductance in  one-dimensional (1D) electron systems
has been studied extensively in connection with the transport
phenomena in quantum wires. It was already established that
the low-energy fixed point of 1D metals is classified as
the Tomonaga-Luttinger (TL) liquid.
The conductance in the TL liquid has been known  
as $2e^2K_{\rho}/h$,\cite{appel,kane}
 where $K_{\rho}$ is the TL 
liquid parameter which controls 
the asymptotic behavior of  correlation functions. 
More recently, it has been found \cite{taru} that 
the experimentally accessible conductance 
may not be $2e^2K_{\rho}/h$ but $2e^2/h$, 
in contrast to the previous theoretical prediction.  
For explaining the 
discrepancy, several possible scenarios have been suggested.  
One proposal is that non-interacting leads attached to 
a quantum wire are essential for reproduction of the observed 
conductance $2e^2/h$.\cite{stone,pome,schu} 
Another proposal claims that if one correctly
takes into account the renormalization of the external
potential due to electron-electron interaction, the factor
$K_{\rho}$ in the conductance may disappear, resulting 
in a desirable expression consistent with
the results of experiments.\cite{kawa} Though either of these 
scenarios seems to explain the 
experimental findings, further extensive studies are 
needed to confirm their relevance to the experiments.  
In what follows, we will calculate the conductance 
based on the self-consistent treatment of the 
external potential.\cite{kawa}

For more detailed analysis of the conductance
in  quantum wires, deviations from the above ideal behavior should
be considered, because the effects
of impurity scattering, Umklapp interaction, etc. 
may become important for real systems of finite
length at finite temperatures. Along these lines,
Ogata and Fukuyama studied the effects of impurity scattering
and long-range Coulomb interaction on the conductance.\cite{fuku}
More recently, Brandes and Kawabata discussed the effect of electron-phonon
interaction.\cite{bran}

If the electron concentration becomes large,
electrons may be affected by a lattice structure and 
then the Umklapp 
interaction is expected to change the behavior of the conductance.
Also, if an artificial lattice structure could be 
fabricated for quantum wires, the Umklapp scattering
should play a crucial role in such a case.  
It is thus interesting to 
investigate the effect of Umklapp scattering on the conductance,
which is the main issue we address in  this paper.
We shall obtain the correction to the 
conductance perturbatively due to  the Umklapp scattering for a 
quantum wire of finite length, 
by taking into account the renormalization of 
the external potential due to electron-electron
interaction.\cite{kawa} We also discuss the temperature dependence of
the  conductance.

Before evaluating the correction to the conductance
explicitly, we first  mention the renormalization of the 
external potential, which was recently
addressed by Kawabata for the TL model.\cite{kawa} He has shown that 
the desirable expression for the conductance,
$2e^2/h$, naturally follows if this renormalization
is taken into account. 
We briefly summarize his argument here.
We consider an  electron system with forward and
Umklapp interactions, 
and start with the following Hamiltonian 
after linearizing the dispersion
around the Fermi points,
\begin{eqnarray}
H&=&
{\rm i}\hbar v_F \int {\rm d}x \sum_{\sigma} :\psi_{\sigma L}^{\dagger}(x)
\partial_x\psi_{\sigma L}(x)-
\psi_{\sigma R}^{\dagger}(x)\partial_x\psi_{\sigma R}(x): \nonumber \\
&+&g\int \frac{{\rm d}x}{2\pi} :\rho(x)\rho(x): \nonumber \\
&+&U\int \frac{{\rm d}x}{2\pi} :{\rm e}^{{\rm i}(4k_{\rm F}-2\pi)}
\psi^{\dagger}_{\uparrow L}(x)\psi_{\uparrow R}(x)
\psi^{\dagger}_{\downarrow L}(x)\psi_{\downarrow R}(x)+{\rm h.c.}:,
\label{hamil1}
\end{eqnarray}
where 
$::$ represents the normal ordering,
$\psi_{\sigma L(R)}$ is the operator for left(right)-moving electrons
with spin $\sigma$,
$g$ ($U$) is the coupling for
forward (Umklapp) scattering, 
and $\rho(x)=\rho_L(x)+\rho_R(x)$,
$\rho_{L,(R)}(x)=\sum_{\sigma}\psi^{\dagger}_{\sigma L(R)}(x)
\psi_{\sigma L(R)}(x)/\sqrt{2}$. 
According to Kawabata,\cite{kawa}
the renormalization of the external potential, $\Phi_0(q,\omega)$, 
occurs due to electron-electron interaction.
In the present case, we have two kinds of 
interactions, both of which may be expected to contribute
to the potential renormalization.
However, as far as the static 
transport properties are concerned,
the renormalization of the potential for large $q$ is irrelevant,
and hence the Umklapp interaction does not play a role in 
this renormalization.
Thus we may consider only the renormalization due to
the forward scattering,
\begin{eqnarray}
-\Phi(q,\omega)&=&-\Phi_0(q,\omega)+\frac{g}{2}
\langle \rho(q,\omega)\rangle \nonumber \\
&=&-\Phi_0(q,\omega)+\frac{g}{2}\chi(q,\omega)\Phi_0(q,\omega).
\label{renp2}
\end{eqnarray}
Here, $\chi(q,\omega)$ is the charge susceptibility.
Then, 
\begin{eqnarray}
\chi(q,\omega)\Phi_0(q,\omega)&=&
\frac{\chi(q,\omega)}{1-\frac{g}{2}\chi(q,\omega)}\Phi(q,\omega) \nonumber \\
&\equiv& \chi'(q,\omega)\Phi(q,\omega).
\label{chip}
\end{eqnarray}
Note that the charge response due to the renormalized potential $\Phi$
is determined by $\chi'(q,\omega)$.
Equation (\ref{chip}) implies that $\chi'(q,\omega)$ is represented by
the diagrams irreducible with respect to the forward interaction $g$,
which can not be separated into two parts by cutting 
any line for forward interaction.
In the absence of the Umklapp interaction,
$\chi'(q,\omega)$ is nothing but the charge susceptibility for
non-interacting electron systems,
and the conductance is reduced to
\begin{equation}
G=2e^2/h, \label{cond}
\end{equation}
as discussed by Kawabata.\cite{kawa}
We show below that including the Umklapp interaction $U$ changes
this expression even at the TL fixed point
where the Umklapp interaction becomes irrelevant
 (see eq.(\ref{condum1})).
We now discuss deviations from the above ideal behavior 
due to the Umklapp scattering.
Using abelian bosonization rules, let us consider the 
standard effective Hamiltonian for eq.(\ref{hamil1}) in terms of 
boson fields.\cite{hal,shan}  
Since  the spin sector is not relevant to the conductance,
we write down here only the charge sector 
in the presence of the Umklapp scattering,\cite{gia}
\begin{eqnarray}
H&=&H_0+H_u, \\
H_0&=&\int {\rm d}x\biggl
[\frac{v_{\rho}}{2 K_{\rho}}
(\partial_x \phi_{\rho}(x))^2+\frac{v_{\rho}
K_{\rho}}{2}(\Pi_{\rho}(x))^2\biggr] \\
H_u&=&\frac{U}{\alpha^2}
\int {\rm d}x \cos(\sqrt{8\pi}\phi_{\rho}(x)+\delta x), \label{umkl}
\end{eqnarray}
where $\alpha$ is the high-energy cut-off parameter.
Here $\phi_{\rho}$ is a boson 
phase field for the charge degrees of freedom,
$\Pi_{\rho}$ is its canonical conjugate field,
and $\delta\equiv 4k_{\rm F}-2\pi$ with
$k_{\rm F}$ being the Fermi point.

As mentioned above, the Umklapp scattering term, eq.(\ref{umkl}), 
becomes irrelevant at the TL liquid fixed point.
Thus for a sufficiently large system at low temperatures, 
the leading correction to the conductance due to
the Umklapp term can be estimated using perturbative calculations.
The conductance is given by
\begin{eqnarray}
G&=&\lim_{\omega\rightarrow 0}
\frac{e^2\omega^2}{L^2}\int^{L/2}_{-L/2} {\rm d}x\int^{L/2}_{-L/2} {\rm d}x' 
\frac{1}{\omega}
\nonumber \\
&\times&
[\langle\phi_{\rho}(x,\omega)\phi_{\rho}(x',\omega)\rangle^R
-\langle\phi_{\rho}(x,0)\phi_{\rho}(x',0)\rangle^R], 
\label{condge}
\end{eqnarray}
where $\langle\cdot\cdot\cdot\rangle^R$ is
the retarded Green's function.
In order to take into account the renormalization of
external potential, we should calculate the irreducible diagram
with respect to the forward scattering, $g$,
which is related to $\chi'(q,\omega)$,
\begin{equation}
q^2\langle\phi_{\rho}(q,\omega)\phi_{\rho}(-q,\omega)\rangle^R_{\rm irr}
=\chi'(q,\omega),
\label{chid}
\end{equation}
for $q\sim 0$. 
$\chi'(q,\omega)$ is expressed in terms of 
$\chi(q,\omega)$ as eq.(\ref{chip}). 
We now separate $\chi(q,\omega)$ into two parts:
$\chi(q,\omega)=\tilde{\chi}(q,\omega)+\delta\chi(q,\omega)$,
where $\tilde{\chi}(q,\omega)$ 
includes only the effect of the forward scattering
and $\delta\chi(q,\omega)$ is the correction due to the Umklapp term.
Then from eq.(\ref{chip}), we have
\begin{eqnarray}
\chi'(q,\omega)&=&\frac{\tilde{\chi}(q,\omega)}
{1-\frac{g}{2}\tilde{\chi}(q,\omega)}
+\frac{\delta\chi(q,\omega)}
{(1-\frac{g}{2}\tilde{\chi}(q,\omega))^2} \nonumber \\
&=&\chi_0(q,\omega)+
\Bigl(\frac{\chi_0(q,\omega)}{\tilde{\chi}(q,\omega)}\Bigr)^2
\delta\chi(q,\omega),
\label{chicor}
\end{eqnarray}
where $\chi_0(q,\omega)$ is the charge susceptibility for non-interacting
electron systems.
To derive the second line, we have used the expression
\begin{equation}
\tilde{\chi}(q,\omega)=\frac{K_\rho v_\rho q^2}
{(v_\rho q)^2-\omega^2},
\end{equation}
 with $K_{\rho}=1/\sqrt{1+g/2v_{\rm F}}$ 
and $K_{\rho}v_{\rho}=v_{\rm F}$,
which hold for the case without the Umklapp scattering.
We now calculate 
$\delta\chi(q,\omega)
=q^2\delta\langle\phi_{\rho}(q,\omega)\phi_{\rho}(-q,\omega)\rangle^R$
up to the second order in $U$.
In order to evaluate this correction, 
we use the relation,
\begin{eqnarray}
&&\langle T \phi_{\rho}(x,\tau)\phi_{\rho}(x',0)
{\rm e}^{{\rm i}\sqrt{8\pi}\phi_{\rho}(x_1,\tau_1)}
{\rm e}^{-{\rm i}\sqrt{8\pi}\phi_{\rho}(x_2,\tau_2)}\rangle_0 \nonumber \\
&=&-\frac{\partial^2}{\partial\alpha\partial\beta}
\langle T {\rm e}^{{\rm i}\alpha\phi_{\rho}(x,\tau)}
{\rm e}^{{\rm i}\beta\phi_{\rho}(x',0)}
{\rm e}^{{\rm i}\sqrt{8\pi}\phi_{\rho}(x_1,\tau_1)}
{\rm e}^{-{\rm i}\sqrt{8\pi}\phi_{\rho}(x_2,\tau_2)}
\rangle_0 \vert_{\alpha=\beta=0}.
\end{eqnarray}
Then, we have
\begin{eqnarray}
&&\delta\langle T\phi_{\rho}(x,\tau)\phi_{\rho}(x',0)\rangle \nonumber \\
&=&\frac{U^2}{2}\int^{L/2}_{-L/2} {\rm d}x_1\int^{L/2}_{-L/2} {\rm d}x_2
\int^{\beta/2}_{-\beta/2} {\rm d}\tau_1\int^{\beta/2}_{-\beta/2} {\rm d}
\tau_2 \nonumber \\
&\times&\{\langle T \phi_{\rho}(x,\tau)\phi_{\rho}(x',0)
{\rm e}^{{\rm i}\sqrt{8\pi}\phi_{\rho}(x_1,\tau_1)}
{\rm e}^{-{\rm i}\sqrt{8\pi}\phi_{\rho}(x_2,\tau_2)}\rangle_0
{\rm e}^{{\rm i}\delta(x_1-x_2)}
+{\rm h.c.}\} \nonumber \\
&=&\frac{U^2}{2}\int^{L/2}_{-L/2} {\rm d}x_1\int^{L/2}_{-L/2} 
{\rm d}x_2
\int^{\beta/2}_{-\beta/2} {\rm d}\tau_1\int^{\beta/2}_{-\beta/2} 
{\rm d}\tau_2 \nonumber \\
&\times&\langle T\phi_{\rho}(x,\tau)\phi_{\rho}(x_1,\tau_1)\rangle_0
\langle T\phi_{\rho}(x_2,\tau_2)\phi_{\rho}(x',0)\rangle_0 \nonumber \\
&\times& \{\langle T {\rm e}^{{\rm i}\sqrt{8\pi}\phi_{\rho}(x_1,\tau_1)}
{\rm e}^{-{\rm i}\sqrt{8\pi}\phi_{\rho}(x_2,\tau_2)}\rangle_0
{\rm e}^{{\rm i}\delta(x_1-x_2)}\nonumber \\
&&+{\rm h.c.}\},
\label{phicor}
\end{eqnarray}
where $\langle \cdot\cdot\cdot \rangle_0$ 
is the average for the gaussian action. 
Thus, from eqs.(\ref{condge}) $\sim$ (\ref{phicor}),
the correction to the conductance is given by 
\begin{eqnarray}
\delta G&=&\lim_{\omega \rightarrow 0}\frac{e^2\omega^2}{L^2}
\int^{L/2}_{-L/2} {\rm d}x\int^{L/2}_{-L/2} {\rm d}x'\int {\rm d}q \nonumber \\
&&\times \frac{1}{\omega}
\Bigl[\Bigl(\frac{\chi_0(q,\omega)}{\tilde{\chi}(q,\omega)}\Bigr)^2
\delta\langle \phi_{\rho}(q,\omega)\phi_{\rho}(-q,\omega)\rangle^R \nonumber \\
&&-\Bigl(\frac{\chi_0(q,0)}{\tilde{\chi}(q,0)}\Bigr)^2
\delta\langle \phi_{\rho}(q,0)\phi_{\rho}(-q,0)\rangle^R
\Bigr]{\rm e}^{{\rm i}q(x-x')}, \nonumber \\
\label{corcon}
\end{eqnarray}
where 
\begin{eqnarray}
&&\delta\langle\phi_{\rho}(q,\omega)\phi_{\rho}(-q,\omega)\rangle^R 
\nonumber \\
&&=\frac{U^2}{2}\int {\rm d}x \int^{L/2}_{-L/2} {\rm d}x_1\int^{L/2}_{-L/2} 
{\rm d}x_2
\nonumber \\ 
&&\times G^R(x,x_1,\omega)G^R(x_2,0,\omega)R^R(x_1,x_2,\omega)
{\rm e}^{-{\rm i}qx} 
\nonumber \\
&&\simeq \frac{U^2}{2}\int {\rm d}x G^R(x,\omega){\rm e}^{-{\rm i}qx}
\int^{L/2}_{-L/2} {\rm d}x'G^R(x',\omega){\rm e}^{-{\rm i}qx'} \nonumber \\
&&\times \int^{L/2}_{-L/2} {\rm d}x''R^R(x'',\omega){\rm e}^{-{\rm i}qx''},
\label{copp}
\end{eqnarray}
with
\begin{eqnarray}
G^R(x-x',\omega)&\equiv&G^R(x,x',\omega) \nonumber \\
&=&\int^{\infty}_0{\rm d}t
\langle [\phi_{\rho}(x,t),\phi_{\rho}(x',0)]\rangle_0
{\rm e}^{{\rm i}\omega t}, \\
R^R(x-x',\omega)&\equiv&R^R(x,x',\omega) \nonumber \\
&=&\int^{\infty}_0{\rm d}t
\{\langle [e^{i\sqrt{8\pi}\phi_{\rho}(x,t)},{\rm e}^{-{\rm i}\sqrt{8\pi}
\phi_{\rho}(x',0)}]\rangle_0 \nonumber \\
&\times&{\rm e}^{{\rm i}\delta(x-x')}+{\rm h.c.}\}{\rm e}^{{\rm i}\omega t}.
\end{eqnarray}

The correction for an infinite system 
at zero temperature is easily obtained by using the renormalization
equations for $K_{\rho}$ and $v_{\rho}$.
The result is
\begin{equation}
G=\frac{2e^2}{h}(1-bU^2)
\label{condum1}
\end{equation}
with $b=K_{\rho}(e^{(4-4K_{\rho})l_c}-1)/(8-8K_{\rho})$.
Here $l_c$ is determined by the condition that 
$|4k_{\rm F}-2\pi|\sim 1/\alpha e^{l_c}$ where $\alpha$ 
is a high-energy cutoff.\cite{gia}
Thus, at the TL fixed point, the deviation from the ideal
conductance due to the Umklapp scattering appears, even if
one takes into account the renormalization of external potential.
However, the prefactor of the conductance is not equal to $K_{\rho}$
in contrast to  a conventional formula for 
the conductance.\cite{appel,kane}

We now consider the correction for finite systems at finite
temperatures.
In order to perform the analytical calculations,
we consider the two limiting cases,  
(i) $L\ll v_{\rho}\beta=v_{\rho}/T$, and (ii)
$L\gg v_{\rho}\beta$.  Case (i) enables us to 
study the system-size effect in 
the low-temperature limit, whereas case (ii)
is suitable for discussions about the temperature
dependence.

\vskip 3mm
\noindent
(i) {\it System-size dependence: $L\ll v_{\rho}\beta=v_{\rho}/T$}

In this case, by  setting $\beta\rightarrow +\infty$, we
can determine the system-size dependence of the conductance at
absolute zero temperature. Then using the conformal transformation
from an infinite complex plane to a strip of width $L$,
\begin{equation}
w=\frac{L}{2\pi}\ln z,
\end{equation}
we obtain the expression for the correlation function, 
$R^R(x-x',\omega)$
for  the finite system of length $L$.
Since it is still difficult to perform the integration
 in eq.(\ref{corcon}) 
in general, we consider two interesting cases, {\it i.e.}
the nearly half-filling case, $\delta L\ll 1$, and the low-density
case, $\delta L\gg 1$, from which we can naturally deduce 
the properties for arbitrary fillings. 

We begin with the case of nearly half-filling, 
 $\delta L\ll 1$. In this case, the Umklapp scattering 
is expected to play a crucial role for a 
system of finite length.
We indeed have the correction,
\begin{eqnarray}
&&\int^{L/2}_{-L/2} {\rm d}x R^R(x,\omega){\rm e}^{-{\rm i}qx} \nonumber \\
&=&c\int^{L/2}_{-L/2}{\rm d}x \int^{\infty}_0 {\rm d}t \nonumber \\
&\times&\Bigl(\frac{\pi}{L}\Bigr)^{4K_{\rho}}
\frac{\cos(\delta x) {\rm e}^{{\rm i}\omega t} 
{\rm e}^{-{\rm i}qx}}{[\sinh\frac{\pi}{L}({\rm i}v_{\rho}t-{\rm i}x)
\sinh\frac{\pi}{L}({\rm i}v_{\rho}t+{\rm i}x)]^{2K_{\rho}}} \nonumber \\
&\sim&\Bigl(\frac{\pi}{L}\Bigr)^{4K_{\rho}-2}c\sin(2\pi K_{\rho})
B(K_{\rho}-\frac{{\rm i}L(\omega+v_{\rho}q)}{4\pi v_{\rho}}, 1 -2K_{\rho}) 
\nonumber \\
&&\times B(K_{\rho}-\frac{{\rm i}L(\omega-v_{\rho}q)}{4\pi v_{\rho}}, 
1 -2K_{\rho})
,\label{r1}
\end{eqnarray}
where $c$ is a non-universal constant, and 
$B(x,y)$ is the beta function.
In the above expression we have 
omitted the constant term independent of $L$,
which was already given in
eq.(\ref{condum1}).
We thus obtain from eqs.(\ref{corcon}), (\ref{copp}), 
(\ref{condum1}) and (\ref{r1}),
the conductance with the correction due to the Umklapp 
scattering,
\begin{eqnarray}
G&=&\frac{2e^2}{h}\Bigl(1-bU^2 
-U^2c\pi\cos^2(\pi K_{\rho})
[B(K_{\rho},1-2K_{\rho})]^2 \nonumber\\
&\times&\Bigl(\frac{\pi}{L}\Bigr)^{4K_{\rho}-3}
\frac{L}{2v_{\rho}}\Bigr).\label{gl1}
\end{eqnarray}
Although the TL liquid parameter $K_{\rho}$
does not appear explicitly in the prefactor of the conductance 
at the fixed point because of the potential renormalization,
it controls the exponent of the system-size dependence
as $\sim L^{4-4K_{\rho}}$ near half-filling. 
Note  that this correction 
is driven by the Umklapp scattering.
The above expression implies that 
the perturbative calculation is justified under 
the condition, $1\gg (U/v_{\rm F})^2(L/\alpha)^{4(1-K_{\rho})}$.
Otherwise, higher order corrections may not be negligible.

On the other hand, in the low density limit, 
 $\delta L\gg 1$,  we have 
\begin{eqnarray}
G&=&\frac{2e^2}{h}\Bigl(1-bU^2 
-U^2c\sin^2(2\pi K_{\rho})
\Gamma^2(1-2K_{\rho})\nonumber \\
&\times&\Bigl(\frac{\delta}{4}\Bigr)^{K_{\rho}-2}
{\rm e}^{-\delta L/2}\frac{L^2}{v_{\rho}}\Bigr).\label{gl2}
\end{eqnarray}
from which one can see that $K_{\rho}$ does not appear in
the exponent of the system-size dependence, similarly to 
the case of free electrons. This
may be naturally understood by recalling that 
 the effect of Umklapp interaction 
becomes small in the low-density limit.
For arbitrary fillings, the finite-size correction to 
the conductance becomes more conspicuous as the 
system approaches half filling.
In any cases away from half-filling, 
the conductance at the TL-liquid fixed point
is reduced to eq.(\ref{condum1}), which has a 
small correction due to the Umklapp scattering.

\vskip 3mm
\noindent
(ii){\it Temperature dependence: $L\gg v_{\rho}\beta$}

We now discuss how the Umklapp interaction affects 
the temperature dependence of the conductance.
To this end, we compute the correction to the
conductance setting $L\rightarrow +\infty$ and
using the following  conformal transformation,
\begin{equation}
w=\frac{v_{\rho}}{2\pi T}\ln z,
\end{equation}
which transforms an infinite complex plane to
a strip whose width is given by the inverse of temperature
$1/T$. For the nearly half-filling case, $\delta v_{\rho}/T\ll 1$,
we thus end up with
\begin{eqnarray}
G&=&\frac{2e^2}{h}\Bigl(1-bU^2
-U^2c\pi\cos^2(\pi K_{\rho})
[B(K_{\rho},1-2K_{\rho})]^2 \nonumber\\
&\times&\Bigl(\frac{\pi T}
{v_{\rho}}\Bigr)^{4K_{\rho}-2}\frac{L}{2T}\Bigr).
\label{compare}
\end{eqnarray}
Therefore $K_{\rho}$ enters the temperature dependence 
of the conductance.
Higher-order corrections may be negligible
provided that $1\gg (U/v_{\rm F})^2(v_{\rho}/T\alpha)^{3-4K_{\rho}}$.
On the other hand, for the low-density case, $\delta v_{\rho}/T\gg 1$,
we have 
\begin{eqnarray}
G&=&\frac{2e^2}{h}\Bigl(1-bU^2
-U^2c\sin^2(2\pi K_{\rho})
\Gamma^2(1-2K_{\rho}) \nonumber \\
&\times&\Bigl(\frac{\delta}{4}\Bigr)^{K_{\rho}-2}
{\rm e}^{-\delta v_{\rho}/2T}\frac{L}{T}\Bigr).
\end{eqnarray}
Here it is seen again  that the TL parameter does not 
appear in the temperature dependence, as mentioned 
before for the system-size dependence.
We note that the above temperature dependence is
 consistent with the results for the 
resistivity at finite temperature obtained by Giamarchi.\cite{gia} 

Through the present analysis of the conductance
for quantum wires, 
it is seen that 
although the TL parameter $K_\rho$ may not be observed 
as a prefactor of the conductance at the TL fixed point, 
it can show up in the 
system-size dependence as well as the temperature dependence
when the Umklapp interaction induces the deviation 
from the TL liquid fixed point.
This is also the case for systems 
with disorder and/or with impurities.\cite{fuku}
If the system contains disorder,  
non-universal temperature dependence including
$K_\rho$ may be observed.\cite{fuku,maslov}  In real materials, 
the effects of disorder as well as the Umklapp interaction
should cause deviations from the ideal TL conductance 
$2e^2/h$.  For example, in the system close to  
half filling,  we have obtained the temperature dependence
(\ref{compare}) driven by Umklapp interaction. However, 
the exponent of the temperature dependence, $3-4K_{\rho}$,
may be larger than that which comes from the 
impurity scattering.\cite{maslov}
Thus in the temperature regime where this behavior
is expected to be observed, disorder in the system may 
control the dominant  temperature dependence. 

In summary, we have investigated the effect of Umklapp interaction
on the conductance in 1D interacting electrons without
disorder. The correction to the conductance due to the Umklapp 
scattering is calculated perturbatively by taking 
into account the renormalization of external 
potential. We believe that the present perturbative approach is
 valid because the forward scatterings are expected to 
predominant over
the Umklapp scattering in conventional quantum wires.
It has been shown that 
the deviation from TL liquid properties manifests itself
in the system-size dependence or the 
temperature dependence of the conductance.
Such a deviation from the ideal TL conductance
becomes more conspicuous for systems close to half-filling.
We expect that quantum wires with an artificial
lattice structure could be fabricated in the near future, for which
the Umklapp interaction should play a crucial role.
Our present findings are expected to be useful in
the analysis of transport phenomena 
in such quantum wire systems.

The authors are grateful to A. M. Tsvelik and T. Brandes for valuable
discussions on the renormalization of the external potential. 
This work was partly supported by a Grant-in-Aid from the Ministry
of Education, Science and Culture.



\end{document}